\documentclass{aa}
\usepackage{graphicx}

\begin{document}

        \title{Envelope models for the supersoft X-ray emission of V1974 Cyg}

        \author{G. Sala \and M. Hernanz}

        \offprints{M. Hernanz}

        \institute{Institut d'Estudis Espacials de Catalunya and Institut de Ci\`encies de l'Espai (CSIC).\\Campus UAB, Facultat de Ci\`encies, Torre C5-parell, 2a planta. E-08193 Bellaterra (Barcelona), Spain.\\ \email{sala@ieec.uab.es, hernanz@ieec.uab.es}}

        \date{Received ... /accepted ...}

        \abstract{The evolution of the soft X-ray emission of V1974 Cyg
          has been simulated by a white dwarf envelope model with steady hydrogen 
          burning. The comparison of the results obtained from ROSAT observations with 
          the results of our envelope models indicates that the post-outburst
          evolution of the nova can be explained by steady H-burning
          on either a 0.9 M$_{\odot }$ white dwarf with 50\% degree 
          of mixing between solar-like accreted material and the 
          ONe degenerate core, or on a 1.0 M$_{\odot}$ ONe white dwarf with 25\% mixing.
}

        \titlerunning{Envelope models for V1974 Cyg}
        \maketitle

%

\section{Introduction}

V1974 Cyg (Nova Cygni 1992), discovered on 1992 February 19, was the first classical
nova to be observed in all wavelengths, from radio to $\gamma$-rays. 
The P Cyg profiles found in all the UV resonance lines of the
emission line spectra observed by IUE (Shore et al. \cite{sho93,sho94}) and 
the strong [Ne III] $\lambda\lambda$3869, 3968 and [NeIV] $\lambda\lambda$3346, 3426
emission lines that dominated the nebular spectrum (Barger et al. \cite{bar93}) 
established V1974 Cyg as a neon nova. V1974 Cyg was observed 
by ROSAT between 1992 April 22 and 1993 December 3 (Krautter et al. \cite{kra96}). 
The X-ray light-curve showed three phases: a first rise phase up
to day 255 after outburst, a plateau phase without great variations
on the flux from day 255 to 511, and a final and fast decline from
day 511 to day 653 after outburst.
During the plateau and decline phases, the X-ray spectrum was dominated
by the soft photospheric emission, which was well fitted with 
MacDonald \& Vennes (\cite{mac91}) ONe enhanced
white dwarf atmosphere models (Balman et al. \cite{bal98}).

Classical nova outbursts are caused by the explosive burning of hydrogen
on the surface of a white dwarf in a cataclysmic variable. 
When a critical amount of H-rich material has been accumulated on the white dwarf surface, 
ignition in degenerate conditions takes place and a thermonuclear runaway is initiated in
the accreted layer. The envelope expands and a fraction of it is ejected at large velocities,
while the rest returns to hydrostatic equilibrium and remains in steady
nuclear burning with constant bolometric luminosity (Starrfield \cite{sta89}). 
As the envelope expands, the photosphere recedes 
and the effective temperature increases, shifting the spectrum from
optical through UV to soft X (MacDonald \cite{mac96}, Krautter \cite{kra02}). 
The soft X-ray emission, arising as the ejecta becomes optically thin
to X-rays, is thus a direct indicator of the thermonuclear burning in the 
post-outburst white dwarf envelope. 
All novae are expected to undergo this phase, showing the
spectrum of a hot white dwarf atmosphere (MacDonald \& Vennes \cite{mac91}),
with effective temperatures in the range  $10^{5}-10^{6}$K and
luminosities close to the Eddington limit. 
The duration of the soft X-ray emitting phase is expected to depend
on the white dwarf mass and the envelope mass left after the outburst.
Without any model of the post-outburst nova, it has been usually estimated 
as the nuclear time-scale of the envelope left, assumed to have a mass similar 
to the accreted layer needed to trigger the outburst, 
$\sim 10^{-4}-10^{-5}$M$_{\odot }$, indicating
nuclear time-scales of tens or hundreds of years 
(Starrfield \cite{sta89}, MacDonald \cite{mac85}).

Nevertheless, X-ray observations indicate much shorter turn-off times for
classical novae. During the last decade, ROSAT observed a total of 39 novae 
less than 10 years after outburst, but only three were found to emit
soft X-rays (Orio et al. \cite{ori01}): 
V1974 Cyg 1992 (with turn-off 18 months after outburst;
Krautter et al. \cite{kra96}), 
GQ Mus 1983 (with turn-off 9 years after outburst;
\"Ogelman et al. \cite{oge93}, Shanley et al. \cite{sha95})
and Nova LMC 1995 (still bright in the year 2000, 
when observed with XMM-Newton; Orio \& Greiner \cite{ori99},
Orio et al. \cite{ori03}). Out of these three novae detected, the 
best observed one was V1974 Cyg, for which the whole evolution of the soft X-ray 
emission was followed.

Here we present a model for post-outburst white dwarf envelope
that can explain the evolution of the soft X-ray emission and turn-off of V1974 Cyg. 
Furthermore, from the comparison of the model with ROSAT observations,
the white dwarf mass, envelope composition and envelope mass of V1974 Cyg are constrained.

\section{White Dwarf Envelope Models for V1974 Cyg}

\begin{table}
\caption{\label{tab1} ROSAT observational results for V1974 Cyg}

\begin{tabular}{ c c c c c }
\hline \hline
\noalign{\smallskip}
 & Day after & K$^{\rm a,b}$ & R$_{\rm photos}^{\rm c}$ & kT$_{\rm eff}^{\rm b}$ \\
 & outburst  & $10^{-25}$              & ($10^9$ cm)              & (eV)\\
\noalign{\smallskip}
\hline
\noalign{\smallskip}
A & 255 & 0.6-2.4 & 1.8-3.7 & 34.3-38.3\\
B & 261 & 0.3-0.9 & 1.3-2.3 & 38.4-41.8\\
C & 291 & 0.4-0.8 & 1.5-2.1 & 41.2-44.3\\
D & 434 & 0.32-0.36 & 1.3-1.4 & 49.4-49.7\\
E & 511 & 0.22-0.26 & 1.1- 1.2 & 50.6-51.0\\
\noalign{\smallskip}
\hline
\noalign{\smallskip}
\end{tabular}
\begin{list}{}{}
\item[$^{\rm {a}}$] Normalization constant of the white dwarf atmosphere
model, $K=(R/D)^{2}$, where R and D are the photospheric radius
and the distance to the source in cm.
\item[$^{\rm {b}}$] Results from Balman et al. (\cite{bal98}).
\item[$^{\rm {c}}$] Photospheric radius for a distance of 2.5 kpc.
\end{list}
\end{table}

A numerical model has been developed to simulate the physical conditions
in the steady hydrogen burning envelope of post-outburst novae. 
A grid of white dwarf envelope models 
has been computed for white dwarf masses from 0.9 to 1.3 M$_{\odot }$.
Three compositions from Jos\'e \& Hernanz (\cite{jh98}) hydrodynamic nova models 
have been considered, corresponding
to ONe novae with different degrees of mixing 
between the solar accreted matter and the degenerate core:
ONe25 models, with 25\% mixing (in mass fractions
X=0.53, Y=0.21, $\delta$X$_{\rm O}$=0.13 -extra O mass fraction 
beyond that in Z-, $\delta$X$_{\rm Ne}$=0.08 -extra Ne mass fraction-; 
Z contains metals in solar fractions); 
ONe50, with 50\% mixing (X=0.35, Y=0.14, $\delta$X$_{\rm O}$=0.26, 
$\delta$X$_{\rm Ne}$=0.16); and ONe75, with 75\% mixing 
(X=0.18, Y=0.08, $\delta$X$_{\rm O}$=0.38, $\delta$X$_{\rm Ne}$=0.24).
Evolution is approximated as a sequence of steady state models 
(for a description of the envelope models see Sala \& Hernanz \cite{sal05}). 

The results show that an envelope with steady H-burning 
proceeds along a plateau of quasi-constant
luminosity, shrinking its photospheric radius as the envelope mass
is reduced, and increasing its effective temperature
as the photosphere sinks into deeper and hotter layers of the envelope.
The average plateau luminosity increases for increasing 
white dwarf masses and for decreasing hydrogen abundances, according to the expression 
$L(L_{\odot})\simeq5.95\times10^{4}\left(\frac{M}{M_{\odot}}-0.536X-0.14\right)$.
Thermonuclear reactions continue until the envelope mass is reduced
down to the minimum critical mass for stable hydrogen burning, 
which occurs shortly after the maximum 
effective temperature is reached. Since no equilibrium configuration
for a smaller envelope mass exists, the shell sources turn-off 
and the white dwarf starts to cool down.
The upper panel in figure \ref{rmt} shows the 
photospheric radius versus the effective temperature for some 
of our models. The lower panels show the
envelope mass for the same models, with the time intervals, in days,
spent by the envelopes to evolve between adjacent points.

For the comparison of our models with V1974 Cyg,  
results from table 1 in Balman et al. (\cite{bal98}) have been used (see table \ref{tab1}). 
For each observation, they obtained a $3\sigma$ confidence range for all model parameters, 
including the effective temperature and the atmosphere normalization constant, 
defined as (R/D)$^2$, where R is the photospheric radius and D is the distance to the 
source. Several determinations of the distance to V1974 Cyg can be found in the 
literature
(Quirrenbach et al. \cite{qui93}, Shore et al. \cite{sho94},
Paresce et al. \cite{par95}, Chochol et al. \cite{cho97}, Balman et al. \cite{bal98}, 
Cassatella et al. \cite{cas04}).
In this work, we adopt the mean value and the whole uncertainty
derived from these distance determinations, 2.5$\pm$0.8 kpc, 
to obtain the photospheric radius from the normalization constants listed 
in table 1 of Balman et al. (\cite{bal98}). 
The contours resulting from this derived radius and the effective temperature range from 
Balman et al. (\cite{bal98}) are overplotted to our models in figure \ref{rmt} (upper panel).
The large uncertainty in the distance
makes an analysis based on the photospheric radius rather unreliable.
Nevertheless, the evolution of the effective temperature alone restricts the
possible models for V1974 Cyg to very few, 
and makes our main results independent from the distance determination.

\subsection{Maximum effective temperature}

A first distance-independent parameter to compare observations and
models is the maximum effective temperature. According to
Balman et al. (\cite{bal98}), the maximum effective temperature observed by ROSAT
was $\sim$50 eV, on day 511 after outburst. 
Nevertheless, the actual maximum temperature
could have been missed by ROSAT, since no observations were
performed between days 511 and 612. 
The V1974 Cyg X-ray light curve (Krautter et al. \cite{kra96}) 
indicates that the end of the
plateau phase, and therefore the maximum effective temperature, occurred
between these two ROSAT observations. Furthermore, 
UV observations indicated that the hot central source had
ceased to photoionize the ejecta about day $\sim$ 530 after outburst 
(Shore et al. \cite{sho96}). In this case, the actual maximum effective
temperature should have been reached before day 530 and thus
should be not much higher than the value obtained for day 511, $\sim$50 eV. 

\begin{figure}
\resizebox{\hsize}{!}{\includegraphics{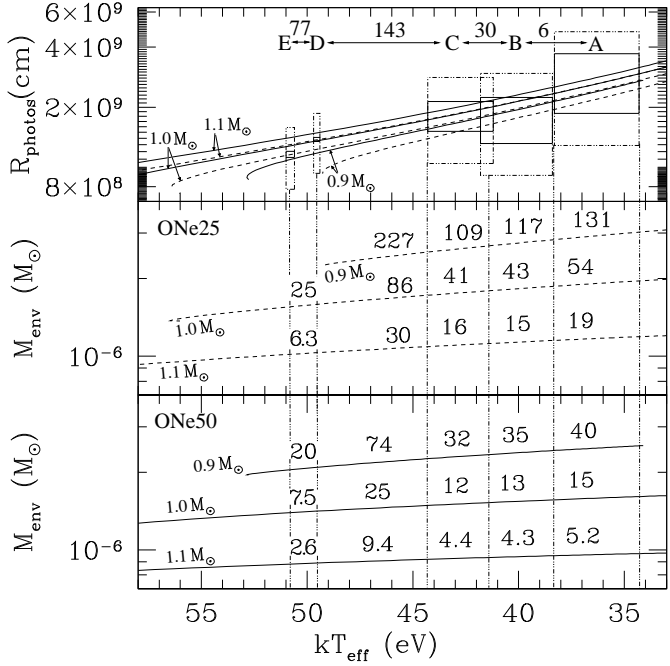}}
\caption{Observational results for V1974 Cyg compared to our envelope models.
\textbf{Upper panel:} Spectral parameters (photospheric radius, R$_{\rm photos}$, 
and effective temperature, kT$_{\rm eff}$) for ROSAT observations 
on days 255(A), 261(B), 291(C), 434(D) and 511(E) after outburst
(from Balman et al. \cite{bal98}. Solid line contours for a distance of 2.5 kpc; 
dotted line contours for all the distance range of 1.7-3.3 kpc).
Observational results are overplotted to ONe25 (dashed) and ONe50 (solid) envelope models, for
white dwarf masses 0.9, 1.0 and 1.1 M$_{\odot}$. Numbers indicate days between observations. 
Vertical lines are orientative for comparison with the lower panels.
\textbf{Middle panel:} Envelope masses for ONe25 models. 
Numbers indicate days spent by the envelope model between two adjacent points.
\textbf{Lower panel:} Same as in middle panel for ONe50 models.}
\label{rmt}
\end{figure}

Taking into account this limit, we find that some envelope models 
are unlikely to represent the observed evolution: for the ONe75 models,
any white dwarf more massive than 0.9  M$_{\odot}$ has
an effective temperature higher than observed. 
For the ONe50 models, we find that only the 0.9 M$_{\odot}$
white dwarf envelopes have a good maximum effective temperature (53
eV, see figure \ref{rmt}). Finally, in the case of the ONe25 composition, this requirement
could be fullfilled by a white dwarf of mass between 0.9  M$_{\odot}$
( kT$_{max}$= 49 eV) and 1.0 M$_{\odot}$ (kT$_{max}$= 56 eV). 
Nevertheless, since evolution at high effective temperatures could proceed faster
than the time interval between observations, this criterium alone is not enough to 
reject models with maximum effective temperatures larger than observed. 
But the evolution of the effective temperature adds some more constrains.

\subsection{Evolution of the effective temperature}

The evolution of the effective temperature provides a second and 
more powerful distance-independent diagnostic tool. 
Only some combinations of core mas and envelope composition can
reproduce the observed evolution.
Interestingly enough, these cases are among the ones with the good 
maximum effective temperature: 0.9 M$_{\odot}$ for ONe50 and 1.0 M$_{\odot}$ for ONe25. 
The days elapsed between ROSAT observations during the
plateau phase are indicated in the upper panel in figure \ref{rmt},
and the corresponding time intervals for the models, in the lower panels. 
Table \ref{tab2} lists the interval times for the models of each composition 
that could better fit the observed evolution.
The time elapsed until day 511 should be taken
with care: if the maximum effective temperature and the subsequent  
turn-off occurred close to that day, several factors can affect the
evolution. It is possible that the X-ray observation took place shortly
after reaching the maximum effective temperature, and not before.
In this case, if the envelope was already starting to cool down, with
still a high effective temperature, the luminosity and effective radius
would have been slightly smaller than the corresponding values for
the source still being on, which could explain the fact that the 
photospheric radius on day 511 is smaller than predcited by models.
The time estimated from the envelope models for the evolution along
the high luminosity branch would then be smaller than observed, and
this is indeed the case for the envelope models with time-scales more
similar to observations.

The ONe75 envelope models (with X$_{H}$=0.18) can not, in any
case, simulate the observed evolution. Even for the smallest white
dwarf in our grid (0.9M$_{\odot}$, which is in fact too small for an ONe white dwarf, 
according to stellar evolution models),
the total time elapsed between effective temperatures corresponding
to days 255 and 511 is $\sim$40 days, smaller than the 256 days elapsed
between observations. For the ONe50 0.9 M$_{\odot}$ white dwarf
envelope models, the time-scale for the same interval of effective
temperatures (141-181 days) is closer to the observed one. Finally, time-scales for 
ONe25 1.0 M$_{\odot}$ models are also similar to the observed ones.
In summary, the best candidates for V1974 Cyg are either 
a 0.9 M$_{\odot}$ white dwarf with 50\% mixing (with X=0.35) 
or a 1.0 M$_{\odot}$ white dwarf with a 25\% mixing (with X=0.53).

In both cases, the envelope mass is in the range $\sim 2\times10^{-6}\rm M_{\odot}$
(see low panels in figure \ref{rmt}) and the luminosity of the model is 
$\sim 3.5\times10^{4}\rm L_{\odot}$. It is worth noticing that, since the 
comparison with models is based on the effective temperature, this 
luminosity determination is independent from distance.

\begin{table}
\caption{\label{tab2} Time intervals observed and simulated by envelope models}

\begin{tabular}{ c c c c c }
\hline \hline
\noalign{\smallskip}
Days between & ONe75 & ONe50 & ONe25 \\
observations & 0.9M$_{\odot}$ & 0.9M$_{\odot}$ & 1.0M$_{\odot}$\\
\noalign{\smallskip}
\hline
\noalign{\smallskip}
256 (A-E)&39.1-43.9&161-201&195-249\\
\hline
179 (A-D)&34.1-38.9&141-181&170-224\\
\hline 
6 (A-B)& $<$13.6& $<$75&$<$97\\
30 (B-C)&$<$17.1&$<$67&$<$84\\
143 (C-D)&17-25.3&74-106&86-127\\
77 (D-E)&5&20&25\\
\hline 
36 (A-C)&8.8-21.9&35-107&43-138\\
173 (B-D)&25.3-34.1&106-141&127-170\\
\noalign{\smallskip}
\hline
\noalign{\smallskip}
\end{tabular}
\end{table}

\section{Discussion}

Among our two candidate envelope models, the 50\% mixing case
is favoured by independent determinations 
of the hydrogen mass fraction in the V1974 Cyg ejecta. 
Austin et al. (\cite{aus96}) found X=0.17 from optical and ultraviolet observations, 
which is similar to the hydrogen abundance in our ONe75 models (X=0.18). 
Nevertheless, as mentioned above, this model would imply a too fast 
evolution compared to ROSAT data. Using mid-infrared spectroscopy, 
Hayward et al. (\cite{hay96}) determined X=0.30, 
very close to the value in our ONe50 models. 
Later works (Moro-Mart\'in et al. \cite{mor01}, 
Vanlandigham et al. \cite{van02}) determined smaller metal enhancements than
Austin et al. (\cite{aus96}), which also agrees with a hydrogen abundance higher 
than X=0.18.

In any of our candidate models, the white dwarf mass 
lays at the lower end of previous determinations.
From their X-ray observations, 
Balman et al. (\cite{bal98}) used the mass-luminosity 
relation from Iben \& Tutukov (\cite{ibe96}) to find a
white dwarf mass in the range 0.9-1.4 M$_{\odot}$. Krautter et al. (\cite{kra96}) estimated 
the star mass to be 1.25 M$_{\odot}$ using the mass-luminosity relation of 
Iben (\cite{ibe82}), which is reproduced by our core-mass luminosity relation above 
for the hydrogen mass fraction of his models, X=0.64.
They took the nova luminosity early in the outburst, $5\times10^4 $L$_{\odot}$,
determined by Shore et al. (\cite{sho93,sho94}), 
who assumed a distance of 3 kpc. Nevertheless,  
later distance determinations situated the nova closer than 
3 kpc and thus the luminosity would be smaller, indicating a less massive 
white dwarf. Moreover, envelope models of Iben (\cite{ibe82}) and 
Iben \& Tutukov (\cite{ibe96}) 
used in both previous mass determinations were hydrogen richer (X=0.64)
than our models, requiring a more massive white dwarf for the same luminosity. 
Our mass determination is in agreement with Retter et al. (\cite{ret97}),
who estimated the mass of the white dwarf to be in the range 
0.75-1.07 M$_{\odot}$ from the periodicities oberved in the light-curve 
and the precessing disc model for the superhump phenomenon. 
A factor cited in previous works 
(Austin et al. \cite{aus96}) in favour of a massive white dwarf 
was the minimum mass for ONe degenerate cores, $\sim1.2$ M$_{\odot}$. 
But recent evolutionary calculations 
in Gil-Pons et al. (\cite{gil03}) have fixed a smaller lower limit, 
showing that final ONe white dwarfs in cataclysmic variables 
have typical masses between 1.0 and 1.1 M$_{\odot}$, thus including
our 1.0 M$_{\odot}$ ONe25 model as a possible one according to 
stellar evolution.

The accreted mass to trigger the outburst of a 1M$_{\odot}$ white dwarf with 
50\% mixing predicted by theoretical models is 
$\sim 6\times10^{-5} \rm M_{\odot}$, whereas the 
ejected mass is $\sim 5\times10^{-5} \rm M_{\odot}$ (Jos\'e \& Hernanz 
\cite{jh98}). Therefore, 
models do not predict remnant envelope masses as low as those with steady 
hydrogen burning that can explain the soft X-ray emission observed for V1974 Cyg 
($\sim 2\times 10^{-6}\rm M_{\odot}$). Since the evolution of V1974 Cyg from day 
255 after outburst to the end of the constant bolometric luminosity phase can 
be explained solely as a result of pure hydrogen burning, there should be some mass-loss 
mechanism able to get rid of most of 
the envelope mass in around 8 months (Tuchman \& Truran \cite{tuc98}), 
but acting at a much lower 
level later on. A mechanism such as a radiation driven wind (Kato 
\& Hachisu \cite{kat94}) 
behaves in the good direction, evolving from large to small rates as envelope mass 
is depleted; however, a fine tuning of various model parameters would be 
required to 
get the particular amount of mass-loss needed. In summary, it is not well 
known which mechanism or mechanisms are responsible for the depletion of the envelope 
mass down to the levels required for the correct interpretation of the X-ray 
emission observed in V1974 Cyg. 

\begin{acknowledgements}
This research has been partially funded by the MCYT project
AYA2004-06290-C02-01  
and by the E.U. FEDER funds. GS acknowledges a FPI grant from the MCYT.
\end{acknowledgements}


\begin{thebibliography}{}
\bibitem[1996]{aus96}Austin, S.J., Wagner, R.M., Starrfield, S., Shore, S.N., Sonneborn, G. \& Bertram, R. 1996, AJ, 111, 2
\bibitem[1998]{bal98}Balman, S., Krautter, J. \& \"Ogelman, H. 1998, ApJ, 499, 395
\bibitem[1993]{bar93}Barger, A.J., Gallagher, J.S., Bjorkman, K.S., Johansen, K.A., \& Nordsieck, K.H. 1993, ApJ, 419, L85
\bibitem[2004]{cas04}Cassatella, A., Lamers, H.J.G.L.M., Rossi, C., Altamore, A. \& Gonzalez-Riestra, R. 2004, A\&A, 420, 571
\bibitem[1997]{cho97}Chochol, D., Grygar, J., Pribulla, T., Komzik, R., Hric, L. \& Elkin, V. 1997, A\&A, 318, 908
\bibitem[2003]{gil03}Gil-Pons, P., Garc\'ia-Berro, E., Jos\'e, J., Hernanz, M. \& Truran, J.W. 2003, A\&A, 407, 1021
\bibitem[1996]{hay96}Hayward, T.L., Saizar, P., Gehrz, R.D., Benjamin, R.A., 
Mason, C.G., Houk, J.R., Miles, J.W., Guli, G.E. \& Schoenwald, J. 1996, ApJ, 469, 854
\bibitem[1982]{ibe82}Iben, I., Jr. 1982, ApJ, 259, 244
\bibitem[1996]{ibe96} Iben, I., Jr. \& Tutukov, A.V. 1996, ApJS, 105, 145
\bibitem[1998]{jh98}Jos\'e, J. \& Hernanz, M. 1998, ApJ, 494, 680
\bibitem[1994]{kat94}Kato, M. \& Hachisu, I. 1994, ApJ, 437, 802
\bibitem[1996]{kra96}Krautter, J., \"Ogelman, H., Starrfield, S., Wichmann, R., \& Pfeffermann.E. 1996, ApJ, 456, 788
\bibitem[2002]{kra02}Krautter, J. 2002, in Classical Nova Explosions, eds. M.Hernanz \& J.Jos\'e, AIP Conference Proceedings, vol. 637, p. 345
\bibitem[1985]{mac85} MacDonald, J., Fujimoto, M.Y. \& Truran, J. 1985, ApJ, 294, 263
\bibitem[1991]{mac91} MacDonald, J. \& Vennes, S. 1991, ApJ, 373, L51
\bibitem[1996]{mac96} MacDonald, J. 1996, in Cataclysmic Variables and Related Objects, Kluwer, Dordrecht, p. 281
\bibitem[2001]{mor01} Moro-Mart\'in, A., Garnavich, P.M. \& Noriega-Crespo, A. 2001, ApJ, 121, 1636
\bibitem[1993]{oge93}\"Ogelman, H., Orio, M., Krautter, J., \& Starrfield, S. 1993, Nature, 361, 331
\bibitem[1999]{ori99}Orio, M. \& Greiner, J. 1999, A\&A, 344, L13
\bibitem[2001]{ori01}Orio, M., Covington, J. \& \"Ogelman, H. 2001, A\&A, 373, 542
\bibitem[2003]{ori03}Orio, M., Hartmann, W., Still, M. \& Greiner, J. 2003, ApJ, 594, 435
\bibitem[1995]{par95}Paresce, F., Livio, M., Hack, W. \& Korista, K. 1995, A\&A, 299, 823
\bibitem[1993]{qui93}Quirrenbach, A., Elias II, N.M., Mozurkewich, D., Armstrong, J. T, Buscher, D.F. \& Hummel, C.A., 1993, AJ, 106, 1118 
\bibitem[1997]{ret97}Retter, A., Leibowitz, E.M., \& Ofek, E.O. 1997, MNRAS, 286, 745
\bibitem[2005]{sal05}Sala, G. \& Hernanz, M., 2005, submitted.
\bibitem[1995]{sha95}Shanley, L., \"Ogelman, H., Gallagher, J.S., Orio, M., \& Krautter, J. 1995, ApJ, 438, L95
\bibitem[1993]{sho93}Shore, S.N., Sonneborn, G., Starrfield, S., Gonzalez-Riestra, R. \& Ake, T.B. 1993, AJ, 106, 2408
\bibitem[1994]{sho94}Shore, S.N., Sonneborn, G., Starrfield, S., Gonzalez-Riestra, R. \& Polidan, R.S. 1994, ApJ, 421, 344
\bibitem[1996]{sho96}Shore, S.N., Starrfield, S. \& Sonneborn, G. 1996, ApJ, 463, L21
\bibitem[1989]{sta89}Starrfield, S. 1989, in Classical Novae, Wiley, New York, p. 39
\bibitem[1998]{tuc98}Tuchman, Y. \& Truran, J.W. 1998, ApJ, 503, 381
\bibitem[2002]{van02}Vanlandingham, K.M., Starrfield, S., Shore, S.N. \& Wagner, R.M. 2002, in  Classical Nova Explosions, eds. M.Hernanz \& J.Jos\'e, AIP Conference Proceedings, vol. 637, p. 224

\end{thebibliography}
\end{document}